# Widely-tunable mid-IR frequency comb source based on difference frequency generation

Axel Ruehl,[1] Alessio Gambetta,[2] Ingmar Hartl,[3] Martin E. Fermann,[3] Kjeld S.E. Eikema,[1] and Marco Marangoni[2,*]

[1]*LaserLaB Amsterdam, VU University Amsterdam, de Boelelaan 1081, 1081 HV Amsterdam, The Netherlands*
[2]*Physics Department of Politecnico di Milano – Polo di Lecco, via Ghislanzoni 24, 23900 Lecco, Italy*
[3]*IMRA America Inc., 1044 Woodridge Avenue, Ann Arbor, MI 48105-9774, USA*
*\*Corresponding author: marco.marangoni@polimi.it*

We report on a mid-infrared frequency comb source of unprecedented tunability covering the entire 3-10 µm molecular fingerprint region. The system is based on difference frequency generation in a GaSe crystal pumped by a 151 MHz Yb:fiber frequency comb. The process was seeded with Raman shifted solitons generated in a highly nonlinear suspended-core fiber with the same source. Average powers up to 1.5 mW were achieved at 4.7 µm wavelength.

Frequency combs have enabled new classes of spectrometers operating with an unprecedented mix of sensitivity, acquisition speed, resolution and accuracy. They are inherently broadband, can be interfaced to high-finesse optical cavities to enhance the sensitivity, and directly referenced to the SI time standard [1]. A major obstacle in the development of a versatile comb-based platform for molecular spectroscopy is the limited tuning range and/or optical power level of the currently available mid-infrared (MIR) frequency combs.

MIR combs can be realized through nonlinear optical processes relying either on optical parametric oscillation (OPO) or difference-frequency generation (DFG). OPOs have been shown to provide W-level frequency combs in the 2.8 - 4.8 µm range [2]. Commercially available nonlinear crystals do not match the requirement for OPO-based systems with a significantly broader tuning range although the pending availability of orientation-patterned GaAs or GaP might change the perspective. Beside the limited tuning range, another drawback is the requirement to lock the OPO cavity to the seeding oscillator which in turn requires two phase-locked loops to obtain phase-stabilization of the MIR output. Self-referencing can indeed be avoided by adopting a degenerate OPO scheme [3], but in this case, the spectral range remains limited to sub-harmonics of the pump wavelength. On the other hand, DFG processes allow for much simpler setups. If pump and signal field are phase-coherent and emitted from the same source, the generated idler field is carrier-envelope-offset phase (CEP) slip-free and requires only stabilization of the comb spacing [4]. DFG-based sources driven by fs-Ti:Sapphire lasers were demonstrated with tuning ranges from 7.5 to 12.5 µm [5]. Er:fiber laser based systems have so far provided tunability either from 3.2 to 4.8 µm [6] or from 5 to 12 µm [7]. The average power levels demonstrated in these approaches were around 100 µW at best. Up to 125 mW from 3 to 4.4 µm has recently been obtained with an Yb:fiber laser by utilizing tunable Raman solitons as seed pulses [8]. No approach has nevertheless been shown to fulfill the requirements for MIR frequency comb spectroscopy in the entire fingerprinting region, where molecular compounds exhibit their strongest absorption features.

In this letter, we demonstrate the use of a low-noise, Yb:fiber frequency comb to produce mW-level MIR pulse trains tunable over an unprecedented range from 3 to 10 µm (1000 – 3350 $cm^{-1}$). The MIR pulses are synthesized by means of a DFG process in a GaSe crystal. A detailed study of the interaction geometry reveals the optimum focusing conditions as a compromise between various constraints hampered by the large tuning range.

The experimental setup is shown in Fig. 1 (a). It is based on a Yb:fiber frequency comb driven by a Fabry-Perot type similariton oscillator operating at 151 MHz comb spacing [9]. Chirped-pulse amplification in a cladding-pumped Yb:fiber produced (after compression) a 2.2 W average power train of < 80 fs pulses centered at 1055 nm [10]. The output was split in two branches, providing tunable seed and fundamental wavelength pump pulses for the subsequent DFG process. As the seed we used the longest wavelength Raman-solitons from supercontinuum generation in a 25 cm long highly nonlinear suspended-core fiber [11]. Fig. 1 (b) shows pump and signal power as a function of the Raman-soliton center wavelength tunable with a slope of 1.8 nm / mW with respect to the launched pump power. Average powers from 1.4 to 1.9 W and about 20 mW for pump and seed pulses, respectively, were available for DFG.

It was recently shown in Ref. [10] that by optimizing fiber and laser parameters, the generated Raman-solitons can be highly coherent. A frequency comb identical to ours was self-referenced and phase-locked to a narrow-linewidth cw-lasers in turn locked to a high-finesse cavity. Out-of-loop comparison between a similar cw-laser at 1.54 µm and the longest wavelength Raman-soliton revealed a linewidth of 1.5 Hz. This corresponds to a coherence length of > 200 ms despite the large Raman-shift of 490 nm. Previous studies on DFG revealed the coherence between pump and signal field as the most critical factor to achieve phase coherence in the MIR [8]. As DFG itself can be realized at an accuracy level of < $10^{-20}$ [4], we expect a high degree of coherence in the MIR.

In our setup, pump and signal beam are combined with a dichroic mirror and focused in a 500 µm long GaSe crystal for Type-I phase-matching (e − o = o). A collinear interaction geometry was chosen to minimize angular dispersion and displacement of the idler beam during tuning. The delay stage in the pump arm enables compensation of the dispersion-induced delay and facilitates temporal overlap. Mode matching was realized by two telescope adjusting the diameter of pump and signal-beam. Behind the GaSe crystal, a Ge long-pass filter separated the residual pump light from the idler output.

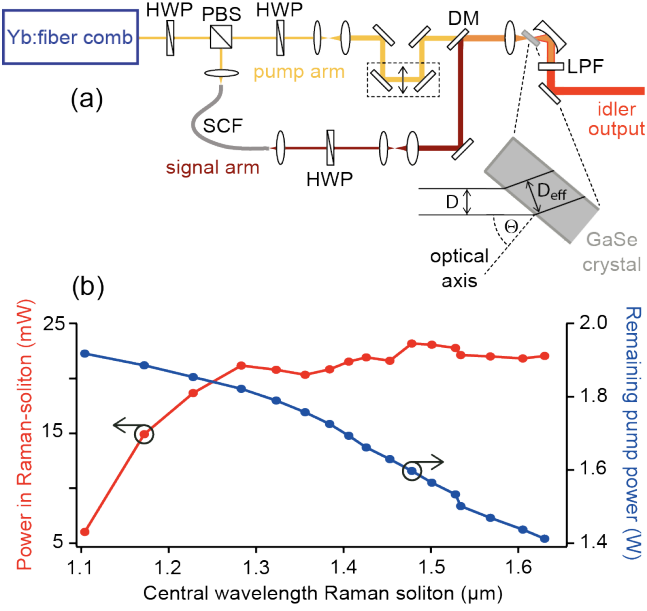

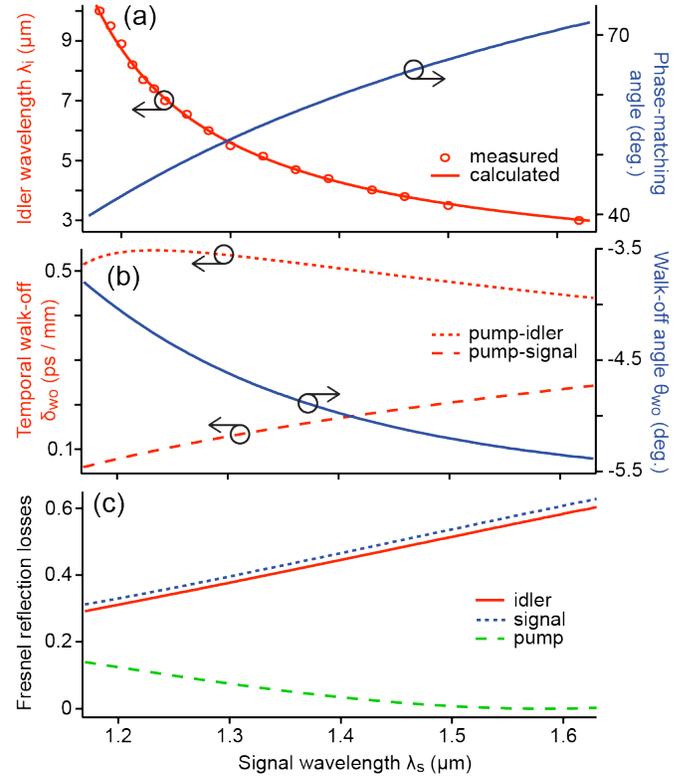

Fig. 1. (Color online) (a) Experimental setup. HWP: half wave-plate; PBS: polarizing beam splitter; SCF: suspended-core fiber; DM: dichroic mirror; LPF: long-pass filter. (b) Average power level of the Raman-solitons (left scale) as a function of its center wavelength together with the residual laser output at 1055 nm (right scale).

The choice of GaSe as a nonlinear crystal was motivated by three main reasons: i) its particularly high birefringence allows angle phase-matching for almost any pair of signal-idler wavelengths (given by $\lambda_i^{-1} = \lambda_p^{-1} - \lambda_s^{-1}$), which is the essential prerequisite for a broadly tunable DFG source (see Fig. 2 (a)); ii) it offers an extremely wide transparency range, from 0.7 to 17 µm; iii) it exhibits a high optical damage threshold and a strong nonlinearity ($\approx 50$ pm / V), 3-4 times higher than periodically poled LiNbO$_3$ (PPLN). As a negative counterpart of these properties: i) the high birefringence implies a strong spatial walk-off between ordinary (signal and idler) and extraordinary (pump) beams (see Fig. 2 (b)); ii) due to the large dispersion at the pump wavelength, a substantial temporal walk-off occurs between pump and signal, and even more severe between pump and idler (see Fig. 2 (b)); iii) GaSe can not be cut at arbitrary angles (only cleaved in the (001) plane), nor can it be AR coated, which translates into strong Fresnel reflection losses at air interfaces due the high refractive index of $n_o \approx 2.8$ and $n_e \approx 2.4$ (see Fig. 2 (c)). Since all these parameters, together with diffraction and dispersion, not shown in the figure, are strongly wavelength dependent, the interaction geometry needs to be chosen as a best compromise between various conflicting constraints.

Fig. 2. (Color online) (a) Measured (dots) and calculated (solid line) idler wavelength (left scale) and external phase-matching angle for Type-I interaction (right scale); (b) Pump-signal and pump-idler temporal walk-off (left scale) and spatial walk-off angle (right scale); (c) Fresnel reflection losses at an air interface. All parameters are plotted as a function of the signal wavelength.

Reducing the beam size increases the conversion efficiency of a parametric process as long as the confocal parameter remains longer than the interaction length. This length is roughly given by the pulse-splitting length $L_{ps}$ between pump and signal, i.e. by the ratio between the pump pulse-width $\tau_p$ (80 fs in our case) and the temporal walk-off $\delta_{wo}$. $L_{ps}$ varies within the tuning range, from 300 to 900 µm when moving towards longer idler wavelengths. Considering an average length of 500 µm, a lower limit to the beam diameter D comes from the spatial walk-off according to the condition $D > D_{wo} = L \cdot \tan \theta_{wo}$, where $\theta_{wo}$ is the spatial walk-off angle shown in Fig. 2 (b). The corresponding $D_{wo}$ suggest optimum beam diameters around 40 µm, but both numerical simulations and experimental results clearly show increasing efficiencies for lower values. This behavior can be understood with the inset of Fig. 1 (a) by observing that the effective beam diameter inside the crystal is magnified in the incidence plane by a factor of M. M approaches its maximum of 1.95 for large $\lambda_i$, so the optimum D value of 18 µm found experimentally is sufficient to overweight the effect of spatial walk-off. At the same time it constitutes a safe choice to prevent possible two-photon absorption and diffraction.

Tuning of the idler output was possible by changing the Raman-soliton wavelength from 1.15 to 1.65 µm with a corresponding adjustment of the temporal overlap and phase-matching angle. The generated MIR spectra recorded with a grating monochromator are shown in Fig. 3 (a). They span the extremely large 3-10 µm range with a spectral width varying between 210 and 710 nm. The maximum power measured with a calibrated pyro-electric detector was 1.5 mW at 4.7 µm corresponding to a peak-power spectral density of about 77 µW / nm and 0.9 µW / comb mode, respectively. The power levels remain in excess of 1 mW in the range from 4 to 5.5 µm, while decreasing to few hundreds of µW at the ends of the tuning range.

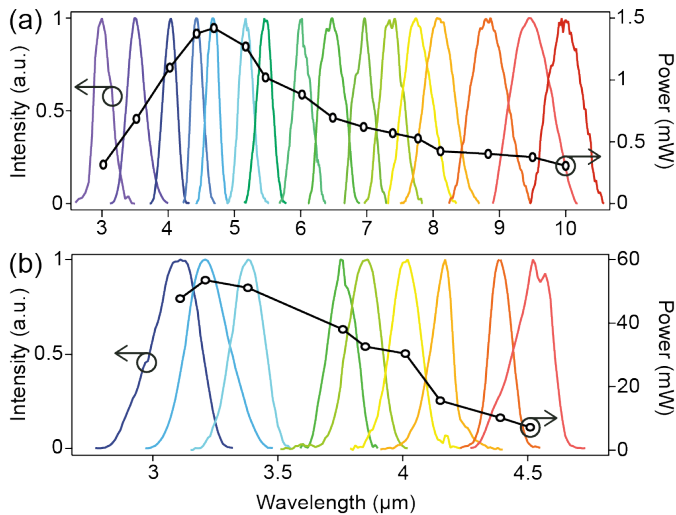

Fig. 3. (Color online) MIR spectra (left scale) and average power (right scale) generated by DFG in (a) a GaSe and (b) a PPLN crystal.

Numerical simulations have been run to evaluate the average power limit of the generated MIR radiation. Some simplifying assumptions were made in the numerical model so it does not take the strong spatial anisotropy of the interaction into account. Nevertheless, the simulations correctly predict the presence of a maximum power around 5 µm, but underestimate the power drop at the wings of the tuning range. On the short wavelength side ($\lambda_i \approx 3$ µm) this is likely to be due to the extreme incidence angles on the crystal (approaching 70°) and the associated phase-front tilting. On the long wavelength side ($\lambda_i \approx 10$ µm) a reasonable explanation is the onset of diffraction not taken into account by the simulations. Experimental power values are on average about a factor of 5 lower than numerical ones. This is to a large extent attributed to the difficulty of keeping the proper spatial overlap and phase-front matching between pump and signal beam over the entire tuning range in particular due to the small spot size of 18 µm.

Significantly higher conversion efficiencies were possible by using a 4 mm long PPLN crystal but with the drawback of a reduced tuning range. As shown in Fig. 3 (b), average powers between 7 and 55 mW were achieved over a tuning range from 3 to 4.6 µm, with the focusing geometry optimized for DFG in GaSe. Extension of the tuning range up to the absorption edge of PPLN at 5.5 µm can be straightforwardly obtained with a different crystal exhibiting a wider range of poling periods.

In summary, a MIR DFG source with a tuning range from 3 to 10 µm at up to 1.5 mW average power using a GaSe crystal is presented. This constitutes one order of magnitude improvement in average power as well as a significant extension of the tuning range compared to previously reported sources based on DFG in GaSe. By using a PPLN crystal, a tuning range from 3 to 4.6 µm at up to 55 mW average power was realized. Since signal and pump fields are generated in the same oscillator, the generated pulses are on average CEP slip-free. Due to the exceptional coherence between the two, we aim to demonstrate the first comb coherence at wavelengths above 5 µm. This will be confirmed with heterodyne beat experiments using a second identical source. In conjunction with the second MIR source, such a system is ideally suited for coherent multi-heterodyne spectroscopy [12] in the molecular fingerprint region.

This work has received funding from the EC's Seventh Framework Program (FP7/2007-2013) under grant agreement n° 228334 and from the Netherlands Organization for Scientific Research (NWO). AR acknowledges personal support from a Marie-Curie-fellowship within the EC's Seventh Framework Program.